\iffalse
\documentclass[runningheads]{llncs}

\def\yesy{yes}
\def\noy{no}
\def\longy{no}
\usepackage{amsfonts}
\usepackage{amssymb}
\usepackage{enumerate}
\usepackage{amsmath}
\def\qed{~$\Box$}
\else
\documentclass[a4paper,UKenglish,12pt]{llncsfp}

\def\yesy{yes}
\def\noy{no}
\def\longy{yes}
\fi
\usepackage{url}
\usepackage{color}
\usepackage[usenames,dvipsnames,svgnames,table]{xcolor}
\usepackage{hyperref}
\hypersetup{colorlinks   = true,
     citecolor    = black,
     urlcolor     = violet,
     linkcolor    = blue}
\def\paraword#1{\medskip\noindent{\bf #1\ }}
\def\paravord#1{\noindent{\bf #1\ }}

\spnewtheorem{thm}{Theorem}{\bf }{\it }
\spnewtheorem{prop}[thm]{Proposition}{\bf }{\it }
\spnewtheorem{prob}[thm]{Open Problem}{\bf }{\it }
\spnewtheorem{cor}[thm]{Corollary}{\bf }{\it }
\spnewtheorem{lem}[thm]{Lemma}{\bf }{\it }
\spnewtheorem{defn}[thm]{Definition}{\bf }{\rm }
\spnewtheorem{rem}[thm]{Remark}{\bf }{\rm }
\spnewtheorem{exmp}[thm]{Example}{\bf }{\rm }
\spnewtheorem{clm}[thm]{Claim}{\bf }{\it }
\spnewtheorem{qust}[thm]{Question}{\bf }{\it }
\spnewtheorem{nota}[thm]{Notation}{\bf }{\rm }

\title{An Exact Algorithm for finding Maximum Induced Matching in Subcubic
Graphs\thanks{Gordon Hoi and Frank Stephan have been supported in part
by the Singapore Ministry
of Education Tier 2 grant AcRF MOE2019-T2-2-121 / R146-000-304-112.
Furthermore, Gordon Hoi was from February to August 2021 employed
by the National University of Singapore.}}
\titlerunning{An Exact Algorithm for Finding Maximum Induced Matching}

\author{Gordon Hoi, Ammar Fathin Sabili\,\inst{1} and
  Frank Stephan\,\inst{1}\,\inst{2}} 
\authorrunning{G.~Hoi, A.F.~Sabili and F.~Stephan}
\institute{School of Computing, National University of Singapore,
13 Computing Drive, Block COM1, Singapore 117417, Republic of Singapore \\
\email{hoickg@gmail.com, ammar@comp.nus.edu.sg}
\and
Department of Mathematics, National University of Singapore,
10 Lower Kent Ridge Road, Block S17,
Singapore 119076, Republic of Singapore \\
\email{fstephan@comp.nus.edu.sg}}

\date{\today}

\begin{document}

\maketitle

\begin{abstract}
\noindent
The Maximum Induced Matching problem asks to find the maximum $k$ such that,
given a graph $G=(V,E)$, can we find a subset of 
vertices $S$ of size $k$ for which every vertices $v$
in the induced graph $G[S]$ has exactly degree $1$.
In this paper, we design an exact algorithm running in $O(1.2630^n)$ time
and polynomial space to solve the Maximum Induced Matching problem for
graphs where each vertex has degree at most 3. Prior work solved the problem
by finding the Maximum Independent Set using polynomial space in the line
graph $L(G^2)$; this method uses $O(1.3139^n)$ time.
\end{abstract}

\section{Introduction}  

\noindent
Most of the graph problems can be classified into either subset, permutation or partition problems.
In this paper, we seek to study a subset problem known as the Maximum Induced Matching problem for subcubic graph 
and we approach it by designing an exact algorithm for it. A more generic
problem of this is known as the Maximum $r$-Regular Induced Subgraph problem.
Given a graph $G=(V,E)$, the Maximum $r$-Regular Induced Subgraph seeks to
find a subset of vertices $S \subseteq V$,
of maximum size such that the induced subgraph $G[S]$ is $r$-regular.
When $r=0$, then this is also known as the 
Maximum Independent Set problem. On the other hand, when $r=1$,
then we have the Maximum Induced Matching (MIM) problem.

MIM has a number of applications in real life. For example, it has
applications in areas such as risk-free marriages \cite{SV82}, 
VLSI design and network flow problems \cite{GL00}. Therefore,
MIM has been studied heavily. It is known that solving MIM
on special instances such as trees and interval graphs \cite{GL00},
chordal graphs \cite{C89}, circular arc graphs \cite{GL93} etc. can be done in
polynomial time. On the other hand, it is also known to be NP-hard in bipartite graphs with maximum degree 4, planar 3-regular graphs,
planar bipartite graphs with degree 2 vertices on one component and
degree 3 vertices in the other component, Hamiltonian graphs,
claw-free graphs, line graphs and regular graphs \cite{DMZ05,KS03,KR03}. 

Gupta, Raman and Saurabh gave a non-trivial algorithm to solve MIM
in $O^*(1.6957^n)$ time, and then improved it to
$O(1.4786^n)$ time in the same paper \cite{GRS12} using polynomial space.
Chang, Hung and Miau also gave an $O(1.4786^n)$ time algorithm to
solve MIM \cite{CHM13}
and using polynomial space as well. Chang, Chen and Hung then gave
an algorithm running in $O^*(1.4658^n)$ time to solve MIM and later improved
it to $O^*(1.4321^n)$ time in the same paper \cite{CCH15}. Furthermore,
Xiao and Tan gave first an algorithm running in $O^*(1.4391^n)$ \cite{XT15}
and then improved in a further paper \cite{XT17}
by giving two algorithms, one that runs
in $O^*(1.4231^n)$ time using polynomial space, the other 
using $O(1.3752^n)$ time and exponential space.

In this paper, we deal with MIM for subcubic graphs (graphs with
maximum degree 3). In the papers mentioned above, most of the
authors deal with low degree graphs by constructing $L(G^2)$
(Line Graph of $G^2$) and then applying an algorithm solving
the Maximum Independent Set problem 
to get the required results. If we were to do so, this allows us to
solve the problem in $O(1.3139^n)$ time
\ifx\longy\noy
(more details in the appendix).
\fi
\ifx\longy\yesy
(more details later).
\fi
Instead, we take a different
approach to design a faster branch and bound algorithm,
running in time $O(1.2630^n)$ time and using polynomial space;
we utilise a variant of the Monien Preis bisection cut method as described
below. Note that this is a polynomial space algorithm; it is known that
there is an algorithm using exponential space which solves the problem
in $O(1.2010^n)$; Kumar and Kumar \cite[Theorems 2 and 4]{KK20} provide
such a type of algorithm and the algorithm for Theorem 2 uses exponential
space and that for Theorem 4 directly applies that of Theorem 2; however,
they maximise the subgraph which contains at least one component of size
$2$ and no component which is larger; this is phrased in terms of
number of nodes removed. The pathwidth result of Fomin and H\o ie
\cite{FH06} can also be used to get the upper bound
of Kumar and Kumar \cite[Theorem 4]{KK20}.

\section{Preliminaries}

\noindent
Given an undirected graph $G=(V,E)$, we let $n$ be the number of nodes $|V|$
and $m$ be the number of edge $|E|$. Given nodes $v,w$ we write $v-w$ for
$v,w$ being the endpoints of an edge and call $w$ a neighbour of $v$.
The degree of a node is the number of its neighbours and the degree of
a graph is the maximum degree of a node in the graph. Graphs of degree
up to three are called subcubic graphs. The line graph of $G$,
denoted as $L(G)$, is constructed from $G$ by having the set of vertices
of $L(G)$ as $E$ and they are adjacent if they are adjacent edges in $G$.
The graph $G^2$ is a graph of $V$, with edges between two vertices
$u$ and $v$ if there is a path of length at most $2$ between $u$
and $v$ in $G$.

A set $B \subseteq E$ of edges is called a {\em bisection cut}
if there are two disjoint subsets $V_1,V_2$ whose union is $V$
such that all edges in $E-B$ have either both endpoints in $V_1$
or both endpoints in $V_2$. The bisection cut is balanced iff
$V_1$ and $V_2$ have each at most $n/2+1$ nodes.

\begin{thm}[Monien and Preis \cite{MP06}] \label{th:monienpreis}
For any $\varepsilon>0$, there is a value $n(\varepsilon)$ such that
the bisection cut of any subcubic graph
$G=(V,E)$ with $|V|>n(\varepsilon)$ is at most
$(\frac{1}{6} + \varepsilon)|V|$.
A possible balanced bisection cut and the corresponding components $V_1,V_2$
can be found in polynomial time.
\end{thm}

\noindent
Note that the above theorem extends to graphs of maximum degree 3 \cite{GS17}.
\ifx\longy\noy
The following corollary is easy to see and has a proof in the appendix.
\fi

\begin{cor} \label{co:monienpreis}
For any $\varepsilon>0$ there is a value $\kappa \geq 3$ such that
whenever the graph has $k > \kappa$ nodes of degree $3$
there is a bisection cut which has on both sides of the cut
have between $k/2-1$ and $k/2+1$ $($inclusively$)$ nodes of degree $3$
and the bisection cut $B$ contains at most $(1/6+\varepsilon) \cdot k$
edges.
\end{cor}

\def\movetotheappendixcomonienpreis{
\paravord{Proof.}
The result is based on the following algorithm which is straight-forward
to verify:

Given a connected graph $(V,E)$, let $V'$ be the set of all nodes of
degree $3$ in $V$ and let $E' = \{(v,w): v,w \in V'$ and $v \neq w$
and there is a sequence of edges connecting $v,w$ such that all nodes
involved other than $v,w$ themselves have degree $2\}$. Such an edge
is called a ``double-edge'' in the case that there are two sequences of the
above type. Note that there are no two neighbouring double-edges
in the graph. Now one applies the Theorem of Monien and Preis for
an $\varepsilon$ chosen below in the algorithm and the corresponding
$\kappa$ to the graph $(V',E')$ and gets a bisection cut $B'$. Whenever
there is a double edge in the bisection cut, this double edge has two
neighbours which are single edges. One moves now one endpoint from
the larger side to the smaller side and so replaces the double
edge by a single edge; the outcome of each such operation is that the
number of nodes of degree $3$ differ at most by $1$ from $k/2$. After
this is done, no edge in the so modified $B'$ is a double edge.
Now one picks for every edge of the modified $B'$ exactly one edge of
the sequence of edges representing the edge in $B'$
and puts this edge it into the new $B$ of the algorithm. The so resulting
$B$ has then at most $(1/6+\varepsilon) \cdot k$ edges.~\qed }

\ifx\longy\yesy
\movetotheappendixcomonienpreis

\begin{rem}
An illustration of this method is the following graph:
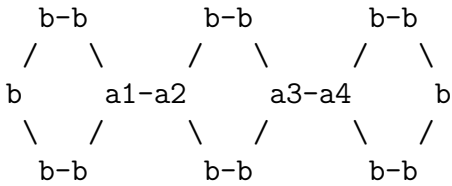
\begin{verbatim}
     b-b       b-b       b-b
    /   \     /   \     /   \
   b     a1-a2     a3-a4     b
    \   /     \   /     \   /
     b-b       b-b       b-b
\end{verbatim}
This graph $(V,E)$ has the following structure $(V',E')$:
\begin{verbatim}
   a1-a2=a3-a4
\end{verbatim}
Here the double edge between a2 and a3 is in the invocation of
the Theorem of Monien and Preis considered a single edge and
$B' = \{a2=a3\}$. Thus one moves the edge one to the side and
then obtains $B=\{a1-a2\}$ splitting only a single edge. Both
halves have $1$ and $3$ nodes of degree $3$, respectively, but
differ only by $1$ from the average of $2$. Here $\kappa$ is too
small to get a value near $3$.

For the choice of $\varepsilon$ and $\kappa$, one chooses
$\varepsilon$ so small and $\kappa$ so large, that the resulting
approximation of the number of degree nodes in each half divided
by the size of the bisection cut is so near to $3$, that using $3$
when computing the branching factors below instead of that approximation
does not lead to different values. Note that all branching factors are
strictly uprounded and therefore one can afford a tiny deviation below
0.0000001 or so from the actual value $3$ without changing the
numerical branching numbers.
\end{rem}
\fi

\ifx\longy\yesy
\medskip
\noindent
We will be using Corollary~\ref{co:monienpreis} in the design of our
algorithm.
\fi

\section{Background on Branching Algorithms}

\noindent
In this section, we will introduce some definitions that we will use
repeatedly in this paper and also the techniques needed
to understand the analysis of the branch and bound algorithm. 
Branch and bound algorithms are recursive in nature and have two kinds
of rules associated with them: Simplification and branching rules.
Simplification rules help us to simplify a problem instance or to act
as a case to terminate the algorithm.
Branching rules on the other hand, help us to solve a problem instance
by recursively solving smaller instances of the problem. 
To help us to better understand the execution of a branch and bound
algorithm, the notion of a search tree is commonly used. 
We can assign the root node of the search tree to be the original
problem, while subsequent child nodes are assigned 
to be the smaller instances of the problem whenever we invoke
a branching rule. For more information of this area,
we refer to the textbooks written by Fomin and Kratsch \cite{FK10}
and by Gaspers \cite{Gas10}.

Let $\mu$ denote our parameter of complexity.
To analyse the runtime of an branch and bound algorithm, one in fact
just needs to bound the number of leaves generated
in the search tree. This is due to the fact that the complexity of
such algorithm is proportional to the number of leaves, 
modulo polynomial factors, that is, $O(poly(|V|,|E|,\mu)$ $\times$
number of leaves in the search tree$)$ $=$
$O^*($number of leaves in the search tree$)$, where the
function $poly(|V|,|E|,\mu)$ is some polynomial based on
$|V|, |E|$ and $\mu$, while $O^*(g(\mu))$ is the class of all
functions $f$ bounded by some polynomial $p(\cdot)$ times $g(\mu)$. 

Now given any branching rule,
let $r\geq2$ be the number of instances generated from this rule
(``outcomes'' or ``alternatives''). Let
$t_1,t_2 ,...,t_r$ be the change of measure for each instance for
a branching rule, then we have the linear recurrence
$T(\mu) \leq T(\mu-t_1) + T(\mu-t_2) + ... + T(\mu-t_r)$.
We can employ 
techniques in \cite{Kul99} to solve it. The number of leaves generated
of this branching rule is therefore 
given as $\beta$, where $\beta$ is the unique positive root
of $x^{-t_1} + x^{-t_2} + ... + x^{-t_r} = 1$. 
For ease of writing, we denote the branching factor of this branching rule
as $\tau(t_1,t_2,...,t_r)=\beta$ and
$(t_1,t_2,...t_r)$ is also known as the branching vector. 

If there are $k$ branching rules in the entire branch and bound
algorithm, with each having a branching factor of
$\beta_1,\beta_2,...,\beta_k$, then the entire algorithm runs
in $O(c^{\mu})$, where $c=max\{\beta_1,\beta_2,...,\beta_k\}$. 

Finally, correctness of branch and bound algorithms usually follows
from the fact that all cases have been covered. 

\def\movetotheappendixcameron{For
Cameron's result \cite{C89}, one has of course to invoke the latest
algorithm and best bound of Maximum Independent Set for a fair comparison.

As mentioned in the Introduction, we can use an algorithm solving the
Maximum Independent Set on $L(G^2)$, where the graph $G$ is a subcubic graph.
The fastest algorithm to solve Maximum Independent Set we are aware
of runs in $O^*(1.1996^n)$ time is by Xiao and Nagamochi \cite{XN17}.

\begin{thm}[Cameron \cite{C89}]
Given a subcubic graph $G$, we can solve MIM for $G$ using an algorithm
to solve Maximum Independent Set in $O(1.3139^n)$ time.
\end{thm}

\paravord{Proof.}
Since $G$ is subcubic, by the hand-shaking lemma, we know that
$m \leq \frac{3}{2}n$. Now we construct the graph
$L(G^2)$, where the edges of $G$ form the nodes of the new graph
and where two nodes of the new graph are connected by a new edge
iff either the two edges share a node or the two edges have each some
node $v,w$, respectively, such that $(v,w)$ is an edge in $G$.
As the number of vertices in $L(G^2)$ is then $m$,
applying Xiao and Nagamochi's algorithm to $L(G^2)$ will then take
$O^*(1.1996^m) \subseteq O^*(1.1996^{\frac{3}{2}n}) \subseteq O(1.3139^n)$
time. Note that the maximum degree of $L(G^2)$ is $12$ and therefore
improved algorithms for Maximum Independent Set at degrees of small
degree do not apply.~\qed
}

\ifx\longy\yesy
\movetotheappendixcameron
\fi

\section{The Algorithm, the Measure and The Rules}

\noindent
This section will give the outline of the algorithm and the
rules. The overall goal is to improve the bound based on the methods
of Cameron \cite{C89} which reduces
Maximum Induced Matching on subcubic graphs to independent set
on line graphs giving complexity $O(1.3139^n)$ to the below
algorithm of $O(1.2630^n)$ time.
\ifx\longy\noy
Please see the appendix for more info on Cameron's result.
\fi

To tighten the analysis of the algorithm, we apply the Measure and Conquer
technique by Fomin, Grandoni and Kratsch \cite{FGK09}. Here, we will be
using a standard weight based measure where we assign 
to each node a weight according to its degree in dependence of a parameter
$s$ with $1/2 \leq s \leq 1$: Nodes of degree $3$ have weight $1$,
nodes of degree $2$ have weight $s$, all other nodes (degrees $0$
and $1$) have weight $0$. The exact value of $s$ will be chosen later.

Let the set of nodes be $V=\{v_1,v_2,...,v_n\}$. We
assign a weight $w_i$ to node $v_i$ as outlined above.
Now we define our measure $\mu = \sum_{i} w_i$. Note that $\mu \leq n$
by definition. Therefore, this means that if our algorithm runs
in time $O^*(c^{\mu})$ for some $c>1$, then we know that our algorithm
runs in time $O^*(c^n)$ because $O^*(c^{\mu}) \subseteq O^*(c^n)$.
The design of such a measure allows us to take
advantage of the fact that whenever  
a node is removed, then each neighbour loses some weight due
to its degrees going down. In particular
a neighbour of degree $2$ loses weight $s$ and a neighbour of degree
$3$ loses at least weight $1-s$.

The overall algorithm consists of the following ingredients: A Monien Preis
subroutine to generate a bisection cut. Simplification Rules which can
be applied at all positions (except of {\bf S4}).
Branching rules {\bf B2.1} and {\bf B2.2}
which have only two possible outcomes (``alternatives'')
and which are applied in the case that one side of
an edge on the bisection cut satisfies the corresponding conditions.
Branching rules {\bf B3.1, B3.2, B3.3} which have $3$ alternatives
and which again require that an edge in the bisection cut is removed
in the process. Branching rule {\bf B4.1} has up to four alternatives
and removes a node being
the end point of two edges in the bisection cut.

The overall algorithm with original input $(V_{inp},E_{inp})$ is recursive
and each recursive step has the following inputs: Current graph $(V,E)$
where $V \subseteq V_{inp}$ and $E = \{(v,w) \in E_{inp}: v,w \in V\}$;
a set $S \subseteq E_{inp}$ of edges which
are already selected for the final maximum subgraph of degree $1$
and all edges $(v,w) \in S$ satisfy that the neither $v,w$ nor their
neighbours in $(V_{inp},E_{inp})$ are in $V$;
current edges $B$ in the bisection cut to be
processed, $B \subseteq E$.
The simplification and branching rules are listed below the algorithm.
We use a recursive algorithm ${\bf AlgoMIM}(V,E,S,B)$ and the initial call is
${\bf AlgoMIM}(V_{inp},E_{inp},\emptyset,\emptyset)$.

\medskip
\noindent
Input: a set of vertices $V$,
a set of edges $E$, a set $S$, a bisection cut $B$. \\
Output: A set $S$ where each node in the induced subgraph $G[S]$ has degree 1.
\\
Function ${\bf AlgoMIM}(V,E,S,B)$ selects the first case of the following
which applies:
\begin{enumerate}
\item If $V=\emptyset$ then return $S$.
\item If $B=\emptyset$ and $V$ consists of $\ell \geq 2$ connected
      components $V_1,V_2,\ldots,V_\ell$ then let $E_h$ be the edges
      of the component given by $V_h$ and
      compute $S_h = {\bf AlgoMIM}(V_h,E_h,\emptyset,\emptyset)$
      for $h=1,2,\ldots,k$. Return the set
      $S \cup S_1 \cup S_2 \cup \ldots \cup S_k$.
\item If $B=\emptyset$ and $(V,E)$ is connected and has at least
      $\kappa+1$ nodes of degree $3$ then one computes
      a new Monien Preis bisection cut $B$ according to the
      bisection cut $B$ given by Corollary~\ref{co:monienpreis} which
      can be computed in polynomial time (as $\varepsilon$ and $\kappa$
      are fixed).
      Let $S' = {\bf AlgoMIM}(V,E,S,B)$ and return $S'$.
\item If one of the simplification rules {\bf S1, S2, S3, S4}
      applies then adjust $V$ and $S$ according to the first applying
      simplification rule and update $E = \{(v,w) \in E: v,w$ are in
      the new $V\}$ and $B = B \cap E$,
      let $S' = {\bf AlgoMIM}(V,E,S,B)$ and return $S'$.
\item If $B \neq \emptyset$ then do the first of the branching rules
      {\bf B2.1, B2.2, B3.1, B3.2, B3.3, B4.1}
      which applies in this order; note that each branching rule in all
      outcomes reduces the size of $B$ by at least one. Now when a
      branching rule produces alternatives $(V_1,E_1,S_1,B_1),\linebreak[3]
      \ldots, \linebreak[3] (V_k,E_k,S_k,B_k)$ then compute for each
      alternative $S_h' = {\bf AlgoMIM}(V_h,E_h,S_h,B_h)$
      and return among these sets $S_h'$ one of maximum size ---
      if there are several maximum ones, it does not matter which one is
      chosen.
\end{enumerate}

\medskip
\noindent
Before giving the rules, it is noted that rules specify several alternatives
and for each alternative they specify the following items: which nodes are
removed from $V$ and which edges are added into $S$, note that the endpoints
of each edge added to $S$ as well as their neighbours are required
to be removed from $V$. $E$ and $B$ are adjusted as indicated in the
above algorithm. For the verification that alternatives cover all possible
cases as well for the calculation of branching numbers, the following two
principles are useful.
\ifx\longy\noy
The verification of these principles and the subsequent simplification rules
is moved to the appendix.
\fi

\paraword{Subset Principle.}
If a branching rule produces (among perhaps others)
the alternatives $(V',E',S',B')$
and $(V'',E'',\linebreak[3] S'',B'')$
to update $(V,E,S,B)$ and if $V-V' \subseteq V-V''$
and $|S''| \leq |S'|$ then one can omit the alternative $(V'',E'',S'',B'')$
from the choices considered.

\def\movetoappendixsubset{
\paraword{Proof-Sketch of Soundness of Subset Principle.}
The reason is that $V'' \subseteq V'$ and $|S''| \leq |S'|$ so that the
output of ${\bf AlgoMIM}(V'',E'',\linebreak[3] S'',B'')$
cannot produce a larger graph than
the output of ${\bf AlgoMIM}(V',E',S',B')$, as at the optimal use the subset
$V''$ of $V'$ does not give more new edges in the output than $V'$
and also the number of old edges in $S''$ is bounded by that of old edges
in $S'$.~\qed }

\ifx\longy\yesy
\movetoappendixsubset
\fi

\medskip
\noindent
An application of the Subset Principle is that
one can first produce a verifiably exhaustive
list of alternatives and then remove some of these with the subset principle.
For simplification rule {\bf S3} all but one alternatives will be removed.

\paraword{Budget Principle.}
If an alternative $(V',E',S',B')$ in a branching rule allows
a subsequent simplification rule then one can assume that this simplification
rule is done directly after the alternative and budget the additional gains
of the measure into the corresponding alternative, giving a better
branching factor.

Technically, one would have to code this in the algorithm (though here
it is only in the verification), this would however make the algorithm
less readable, as one would need more cases to expand this into one
reasoning. The budget principle is only applied in few subcases of the
verification and calculation of $\tau$-numbers of the three-way branching rules.

\paraword{On Diagrams and Conventions.}
In the subsequent text, nodes $a,a',a'',\ldots$ have always degree $3$,
nodes $b,b',b'',\ldots$ have degree $2$, nodes $c,c',c'',\ldots$
have degree $1$, nodes $d,d',d'',\ldots$ have either degree $2$
or degree $3$ and nodes $e,e',e'',\ldots$ are not apriori specified.
Furthermore, $a-b$ denotes that there is an edge from $a$ to $b$
and $a-|-b$ denotes that there is an edge from $a$ to $b$ which
is part of $B$.

\paraword{Simplification Rules.}
Simplification rules {\bf S1} and {\bf S2}
handle isolated components with $\kappa$
or less nodes of degree $3$, where $\kappa$ is a suitably chosen constant.
Simplification rule {\bf S3}
removes various special cases which do not need to be considered
\ifx\longy\noy
in branching rules, see the appendix for examples.
\fi
\ifx\longy\yesy
in branching rules, for examples see after the rule below.
\fi
Branching rule {\bf S4} abstains from branching an edge as shown below
but instead moves the node $c$ onto the other side
of the bisection cut and so {\bf S4} avoids a bad case in branching.
\pagebreak[3]

\begin{verbatim}
S1. e   c-c   c-b-c  c-b-b-c   b-b-..-b-b   All nodes
                               |        |   have degree
    c-b-b-b-c   c-b-..-b-c     b-b-..-b-b   0, 1 or 2.

S2. Constantly many nodes of degree 3

S3.                c           b        a-c       a-c
                  /           /|       /|        /|
..-d-b-c    ..-d-a-c    ..-d-a-b   ..-d-b    ..-d-a-c
\end{verbatim}
\pagebreak[3]
\begin{verbatim}
S4. ..-d-|-c  (c is a node of measure 0 on other side of
               bisection cut)
\end{verbatim}

\paravord{Simplification Rule S1.} If there is a component without nodes of
degree 3, then simple computations show that degree 0 nodes cannot
contribute to $S$, an edge of the form c-c goes straight into $S$
and a line of the form ending with c having $3k+1, 3k+2, 3k+3$ edges
contributes $k+1$ edges to $S$ (by having a pause of two edges between
any two edges going into $S$) and a cycle of $3k+3, 3k+4, 3k+5$ edges
contributes $k+1$ edges (again by having a pause of at least two edges
between any two edges going into $S$).

\paraword{Simplification Rule S2.} If there are at most $\kappa$
nodes of degree $3$, then one can compute in time $3^\kappa \cdot Poly(n)$
an optimal $S'$ restricted to this component and then one replaces $S$
by $S \cup S'$ and removes the component from $(V,E)$.

\paraword{Simplification Rule S3.}
\def\sthree{Assume that $d$ is a node,
$D$ a proper nonempty subset of the set
of its neighbours and that $C$ be the nodes outside $\{d\} \cup D$
which are neighbours of the nodes in $D$. If there is at least one
edge with both endpoints in $C \cup D$ and if all edges with an endpoint
in $C \cup D$ has one end in $D$ and the other one in $C \cup D \cup \{d\}$
then one puts into $S$ a maximal and legal set of edges with both
endpoints in $C \cup D$ and removes $\{d\} \cup C \cup D$ from $V$.}
\sthree
\ifx\longy\noy
See the appendix for some examples.
\fi

\paraword{Simplification Rule S4.} If due to removal of other nodes it
happens that there is an edge in $B$ which connects a degree 1 node
with a node on the other side of the bisection cut, then move the
degree 1 node to the other side and remove the edge from $B$.

\medskip
\noindent
The verification of the simplification rule {\bf S1} is standard and omitted.
\ifx\longy\noy
A polynomial time algorithm for {\bf S2} and the proofs of
soundness of {\bf S3} and {\bf S4} are moved to the appendix.
\fi

\def\movetotheappendixsimplification{\paraword{Possible
Polynomial Time Algorithm to Implement S2.}
As long as there is a degree $3$ node $a$, one picks a neighbour $e$
and branches with $a-e$ being an edge to go into $S$ versus only 
$a$ being removed from the graph versus only $e$ being removed from the graph;
in the first case, $a-e$ goes into $S$ and all neighbours of $a$ and
$e$ are removed together with these nodes. This algorithm processes
$\kappa$ three-way branchings giving a $3^\kappa$-sized tree such that
each leave consists of up to $3\kappa$ components without nodes of
degree $3$ which can be treated by using simplification rule {\bf S1}.
The algorithm then takes the maximum $S'$ produced by any of these
$3^\kappa$ alternatives.

\paraword{Proof of Soundness of Simplification Rule S3.}
Note that if $d$ is endpoint of an edge selected for $S$ then all nodes
in $D$ are removed (and perhaps some in $C$) and no edge remains with
an endpoint in $C$ and therefore no further edge can
be put into $S$ than the one mentioned; on the other hand, there is an
edge between two nodes of $D \cup C$ and thus if one selects this edge for
$S$ then only the nodes in $C \cup D \cup \{d\}$ have to be removed. Thus
by the subset principle, one puts this edge into $S$ and removes the
set $C \cup D \cup \{d\}$ from $V$.

Note that in some few situations one
can put two edges into $S$, but then $d$ is also not part of any of
these edges. The graphical enumeration in {\bf S3} is not covering
all arising cases.~\qed 

\paraword{Proof-Sketch of Soundness of S4.}
Note that the invariance is that on both
sides of the bisection cut the number of degree 3
nodes differs at most by $2$
at the start. The moving over of a degree 1 node
does not destroy this balance, but reduces the duty to branch by
one edge.~\qed }

\ifx\longy\yesy
\movetotheappendixsimplification

\paraword{Examples for Simplification Rule S3.}
Assume the following four situations:
\begin{verbatim}
  (1) e-d-b-c    (2) e-d-a-c   (3) e-d-a-d'-b'-c'  (4) e-d-a'-c'
                       |  \            | |               | |
                       e'  c'          c e'            c-a-b
\end{verbatim}
The nodes $d,d'$ are those which have in the rule {\bf S3}
the name $d$. After applications of {\bf S3} only nodes $e,e'$
remain from the above displayed parts of the graph.

In the first situation, $D = \{b\}$ and $C = \{c\}$.
The edge $b-c$ goes into $S$ and $d,b,c$ are removed.

In the second situation, $D = \{a\}$ and $C = \{c,c'\}$.
One of the edges $a-c$, $a-c'$ go into $S$
and nodes $a,c,c',d$ are all removed.

In the third situation, there are two subsequent applications
of {\bf S3}. The immediate application uses $d'$, $D =\{b'\}$ and
$C = \{c'\}$ and puts $b'-c'$ into $S$ and removes $b',c',d'$ which
makes $a$ to become a degree $2$ node.  The next application uses $d$
has $D = \{a\}$ and $C = \{c\}$, puts $a-c$ into $S$ and
removes $a,c,d$.

The fourth situation has $D = \{a,a'\}$ and $C = \{b,c,c'\}$ and the edges
$a-c, \linebreak[3]  a'-c'$ are put into $S$. The nodes $a,a',b,c,c',d$
are all removed from the graph.
\fi

\paraword{Branching Rules.} All branching rules will be set up such
that in all alternatives at least one edge of the bisection cut is
removed. Here the bisection cut is computed from Corollary~\ref{co:monienpreis},
not from the original theorem. For accounting purposes,
$V'$ is the set of degree $3$ nodes in the current instance, then
the bisection cut $B$ is set up in such a way that it contains
$|V'|/6+o(|V'|)$ edges
and splits the graph into two halves such that each of them
having at most $|V'|/2+1$ degree $3$ nodes. Now the average
$|V'|/2 / |B|$ tends to $3$ for arbitrary large $V'$.
Thus the constants $\varepsilon,\kappa$ are selected such that the difference
between $3$ and $(|V'|/2-1)/|B|$ is so small that computing the branching
factors with the approximation $3$ or the latter constant does not influence
the uprounded numerical value to four decimal places.
In particular we analyse the branching factors from each side of
the bisection cut separately and take into account the cut-off
nodes as resolved. More precisely,
for measure gains on the other side of the bisection cut, one only accounts
the constant $3$, for the measure gains on the own side of the bisection
cut, one computes according to the nodes removed or losing weight.

The branching rules are in the following situations where
in the subsequent diagrams $-|-$ denotes the edge in the bisection cut.
The other edges may or may not be in the
bisection cut, it does not matter, as only the removal of one edge
from the bisection cut in the two-way branching gives enough measure.

\paraword{Two-Way Branching Rules.}
In the following three situations, one can either apply a simplification
rule or resort to two-way branching which allows for good branching factors.

\begin{verbatim}
S4 (see above)   B2.1           B2.2

d-a-|-c          d-|-a-d'       d-|-a-d'
  |                  |              |/
  d'                 c              b
\end{verbatim}

\paravord{Branching Rule B2.1.}
One removes $a,c$ from $V$ versus putting $a-c$ into $S$ as an edge and
removing $a,c,d,d'$ from $V$.

\paraword{Branching Rule B2.2.}
One removes $a$ from $V$ versus putting the edge $a-b$ into $S$ and removing
$d,a,b,d'$ all from $V$.

\paraword{Explanation and Verification.}
Note that the actual case-distinction in {\bf B2.1}
is removing $a$ without putting an edge into $S$
versus putting an edge with $a$ into $S$;
when $a$ is removed without putting an edge into $S$,
then the node $c$ becomes a degree $0$ node and can
also be removed in the same action. Note that $d,d'$ have at
least one additional neighbour of degree $2$, as otherwise
simplification rule {\bf S3} would apply.

The subset principle applies in all cases for completeness. The
reason is that for node $c$ in {\bf B2.1} as well
as node $b$ in {\bf B2.2} satisfy that all their neighbours are also
either the respective node $a$ in the corresponding rule
{\bf B2.1, B2.2} or its neighbours.
As every edge in which the main node $a$ is involved
requires the removal of $a$ and all its neighbours, the corresponding
removed parts of $V$ include always the part which is taken when one puts into
$S$ the edges $a-c$ / $a-b$ in the respective rules. Thus in both
cases one has two-way branching and each time the two-way branching removes
on both sides an endpoint of the edge in the bisection cut and thus the edge
in the bisection cut itself. Therefore one can always put measure $3$ into the
budget for removing an edge from the bisection cut and furthermore some
nodes are removed or lose weight. However, in the case that both have
degree $2$, the loss of measure is higher, as simplification rule {\bf S3}
applies after removing $a,c$ only; therefore one can assume for the worse
case that the neighbour of $d$ has degree $3$.
On the side of $d$ the weight-loss is smaller, as $d$ and its neighbour
might both be degree $2$ nodes. If $d$ has degree $2$ the branching factor
is $\tau(3+s,4)$ and if $d$ has degree $3$ the branching factor is
$\tau(4-s,5-s)$, in both cases assuming that the neighbour has degree $3$
for lower weight loss when downgrading the degree.

\paraword{Three-Way Branching Rules.}
The situations of the rules follow the below graphics;
only edges between nodes of
both sides are listed. In the following diagrams, the edge \verb:-|-:
denotes an edge in the bisection while normal lines are edges not in the
bisection cut.
\pagebreak[3]
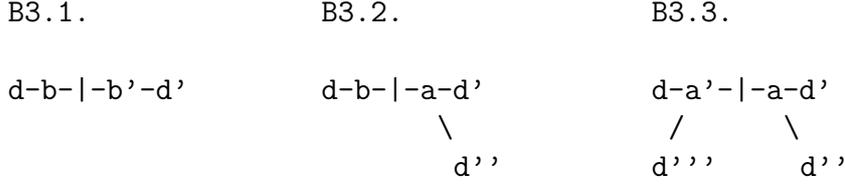
\begin{samepage}
\begin{verbatim}
    B3.1.              B3.2.               B3.3.
    
    d-b-|-b'-d'        d-b-|-a-d'          d-a'-|-a-d'
                              \             /      \
                               d''         d'''     d''
\end{verbatim}
\end{samepage}

\paravord{Three-Way Branching Rule B3.1.} The endpoints $b,b'$ of the edge
in the bisection cut have degree $2$ and all their neighbours have either
degree $2$ or degree $3$. In each of the $3$ branches, an edge goes into
$S$, namely $d-b$ or $b-b'$ or $b'-d'$; the endpoints of the edges and
their neighbours will be removed from $V$ and $E$ will be updated accordingly.

\paraword{Three-Way Branching Rule B3.2.} One branches $b$ into $S$ versus
$b$ removed as single node; thus the $3$ options are to put edge $d-b$
into $S$; to put the edge $b-a$ into $S$; just to remove node $b$ from $V$.
When putting an edge into $S$, its endpoints and their neighbours are removed
from~$V$.

\paraword{Three-Way Branching Rule B3.3.} One either puts the edge
$a'-a$ into $S$ and removes $a,a',d,d',d'',d'''$ from $V$ or one
removes only $a'$ from $V$ or only $a$ from~$V$.

\paraword{Coverage of all cases by the various rules.} For rules {\bf B3.2} and
{\bf B3.3} it is clear that the case distinction is complete: In the case of the
rule {\bf B3.2}, the case distinction is that no edge goes into $S$ and
just $b$ is removed versus the two cases where an edge with endpoint $b$
goes into $S$; in the case of
rule {\bf B3.3}, the case distinction is that the edge $a'-a$ goes into $S$
versus the two cases that either $a$ or $a'$ are removed from $V$ without any
edge going into $S$. For the rule {\bf B3.1}, more work is required.
 
So assume that one looks whether the nodes $d,d'$ are part
of an edge which goes into $S$ or not. If they are not, then the two nodes
between them can go into $S$, that is the case that the edge selected is
$b-b'$. If one of them is part of an edge going into $S$
but the other one not, then for
this one edge the nodes $b,b'$ between $d,d'$ can only be part of the
edge of the border node or not part of any edge going into $S$.
Thus, by the subset principle, these cases are supersets of case that
the selected edge is $b-b'$.
The last case where edges of $d,d'$ go into $S$. If both edges do not
have endpoints $b,b'$, respectively, then there is a proper subset having
two edges going into $S$ which could also be removed from $V$, thus
this case does not apply. So one of the edges is $d-b$ or $b'-d'$,
respectively. In these cases, instead of taking out the full set, one
just considers the subsets where $d-b$ or $b'-d'$ go alone to $S'$
and removes the other side $d'$ and $b$, respectively, from the nodes
to be removed from $V$ to simplify the subcases.
Thus the case-distinction from rule {\bf B3.1} is indeed legitimate.

\paraword{Branching Numbers of Three-Way Branchings B3.1 and b-side of B3.2.}
A detailed analysis of the branching cases shows that for {\bf B3.1}
and for the $b$-side of {\bf B3.2} the $3$ cases do the following:
the first one removes just the node $b$ (that is selecting the edge
$b'-d'$ in {\bf B3.1} and just removing $b$ in {\bf B3.2}); the next
one removes the nodes $d,b$ which is selecting the edge $b-b'$ for
$S$ in {\bf B3.1} and selecting the edge $b-a$ in {\bf B3.2}; the
third one removes the nodes $d,b$ as well as all not listed neighbours of
$d$ on the $b$-side from the bisection cut. There are several cases how
the side $d-b$ looks like; in cases (a) and (b), $d$ has degree $2$;
in the cases (c), (d), (e) and (f), $d$ has degree $3$.
\pagebreak[3]
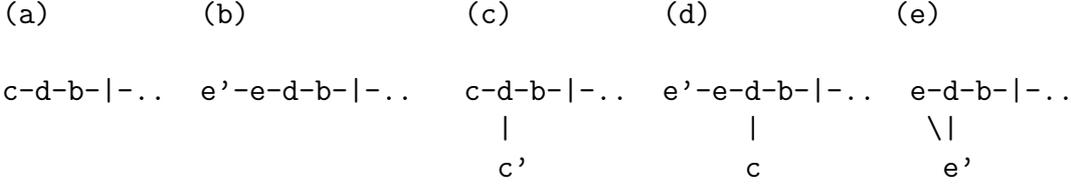
\begin{samepage}
\begin{verbatim}
(a)         (b)             (c)         (d)           (e)

c-d-b-|-..  e'-e-d-b-|-..   c-d-b-|-..  e'-e-d-b-|-..  e-d-b-|-..
                              |              |          \|
                              c'             c           e'

(f) Same as (e), but e,e' do not have a joint edge.
\end{verbatim}
\end{samepage}
Case (a) and case (c) do not occur, as simplification rule
{\bf S3} covers these cases and the branching rules are
not reached. Similarly, if $e'$ would have degree $1$
and any further neighbours of $b,d,e$ (if any) would have
degree $1$ then simplification rule {\bf S3} would remove
some nodes prior to branching and therefore also this case
does not apply.

In the following it is shown that, using $1/2 \leq s \leq 1$,
the branching number is always at least as good as
$\max\{\tau(3+4s,3+4s,3+4s),\tau(4,4+s,5+s)\}$.

In cases (d), (e), (f), if one removes $b$ then this gives
weight $s$ and $d$ is losing weight $1-s$, together with the
constant $3$ for removing an edge from $B$, it gives $4$.
If one removes $d,b$ then the gain is at least $4+s$ without
taking weight loss of neighbours into account.
If $e$ or $e'$ has weight $1$ then the gain is at least $5+s$.
If $e$ and $e'$ have both weight $s$ then the gain is $4+3s$
which is above $5+s$ since $s \geq 1/2$.

So it remains the case (b). If $e$ has degree $3$ then the
weight gain at removing $b$ is $3+2s$, the weight gain at
removing $d,b$ is $3+2s+1-s = 4+s$, the weight gain at
removing $d,b,e$ is in the worst case $3+(1-s)+2s = 5+s$.
As $s \geq 1/2$, $\tau(3+2s,4+s,5+s)$ is also better than
$\tau(4,4+s,5+s)$. If $b,d,e$ all have degree $2$ then
removing $b$ also removes $d,e,e'$ and so one gets
either $\tau(4+3s,3+3s,4+2s)$ or, if all four have degree $2$,
then one can apply simplification rule {\bf S3} both after
removing $b$ which takes away $d,e,e'$ and after removing
$b,d$ which takes away $e,e'$ and a neighbour of $e'$.
The third case is that $b,d,e$ are removed and $e'$ goes
down to weight $0$. Note that $4 \leq 3+3s$, $4+s \leq 4+2s$
and $5+s \leq 4+3s$, thus $\tau(4+3s,3+3s,4+2s)$ is better
than $\tau(4,4+s,5+s)$. In the other case, the 
branching factor is better than $\tau(3+4s,3+4s,3+4s)$,
as all four nodes are by follow-up branchings eliminated
in all three alternatives of the branching.

\paraword{Branching Numbers of Three-Way Branchings a-side of B3.2 and B3.3.}
Note that the following $3$ alternatives occur on
the current side of the bisection: only the edge in the bisection cut is
removed; the node $a$ is removed; the nodes $a,d',d''$ are removed.
The first case arises when for the rule {\bf B3.2} only $d$ is removed
and when for the rule {\bf B3.3}, $a$-side only the node $a'$ on the
other side is removed. The next case occurs at {\bf B3.2} when the edge
$d-b$ is selected for $S$ and thus the neighbour $a$ is removed from $V$
and for {\bf B3.3} when the node $a$ is only removed. The third case,
removal of $a,d',d''$ on the current side of the bisection cut occurs
in rule {\bf B3.2} when the edge $b-a$ is selected for $S$ and thus
also the nodes $d',d''$ are removed and in the rule {\bf B3.3} when
the edge $a'-a$ is selected for $S$ and thus also the neighbours $d',d''$
are removed from $V$. For evaluating the branching number, one again
has to distinguish several configurations. Cases (a), (b) and (c) -- unless
$d,d'$ in (c) have both degree $3$ and one of them has a further
neighbour of degree $2$ -- do not
occur, as the corresponding simplification rules {\bf S3}
and branching rules {\bf B2.1} and {\bf B2.2} apply.
\pagebreak[3]

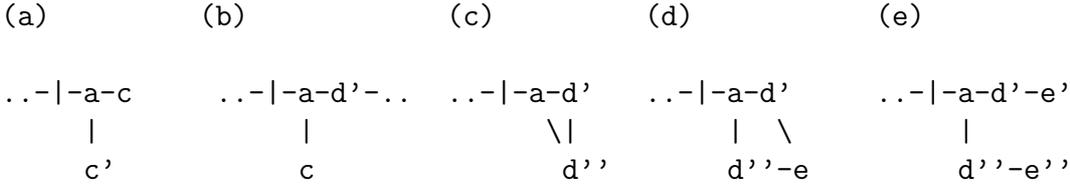
\begin{samepage}
\begin{verbatim}
(a)         (b)            (c)         (d)           (e)

..-|-a-c     ..-|-a-d'-..  ..-|-a-d'   ..-|-a-d'     ..-|-a-d'-e'
     |            |              \|         |  \          |
     c'           c               d''       d''-e         d''-e''
\end{verbatim}
\end{samepage}
\pagebreak[3]

\noindent
For these graphics, $d',d''$ have been replaced by
$c,c'$ when their degree is $1$ in cases (a) and (b); this only occurs
in cases captured by prior simplification or two-way branching rules.

For (c), if both $d',d''$ have degree $3$ and both have an additional
neighbour of at least degree $2$ then by the above mentioned
exception, neither branching rule {\bf B2.2} nor simplification
rule {\bf S3} do apply, but the branching
according to {\bf B3.2} or {\bf B3.3} gives
either $\tau(4-s,6-2s,8-2s)$ or $\tau(4-s,6-2s,6+s)$
where the weight-loss of the further neighbours of $d',d''$ in the
last alternative is in the worst case is either $s$ (if neighbours identical)
or $1-s$ each (if distinct neighbours).
If one of $d,d'$ has only an additional neighbour of degree $1$ and
the other one has an additional neighbour of degree $2$ or $3$ then
simplification rule {\bf S3} applies in the second alternative when
removing $a$ so that furthermore $d,d'$ are removed and the worst case
branching factor is $\tau(4-s,7-s,7-s)$. The case that both have only
an additional neighbour of degree $1$ is eliminated by simplification
rule {\bf S3} prior to branching.

In (d) one distinguishes several cases. First, the case that
all nodes $d',d''$ have degree $2$. Then simplification
rule {\bf S1} (if $e$ has also degree $2$) or simplification rule {\bf S3}
(if $e$ has degree $3$ and is connected to something) can apply and the
corresponding gain of measure allows to evaluate the subcase
as $\tau(4-s,5+s,5+2s)$. 
Second, it does not happen that $e$ has degree $2$ and $d',d''$ have
besides $a,e$ only degree $1$ neighbours (if any), as then simplification
rule {\bf S3} would have removed this part prior to branching.
Third, the case that $d',e$ have degree $3$ and $d'$ has a neighbour
$c$ of degree $1$. Then when removing only $a$, the simplification rule
{\bf S3} removes $c,d',e$ which gives the corresponding gain of additional
measure and $\tau(4-s,6+s,6+s)$. 
Fourth, the case that $d'$ has a neighbour of degree at least $2$ and
$e$ has degree $3$. Then the case that $a,d',d''$ are all removed
reduces the weight of $e$ by $1$ and brings in a further loss of
measure $1-s$ for the additional neighbour of $d'$, resulting in
the branching factor $\tau(4-s,5,7-s)$.
Fifth, the case that $d',d''$ both have besides $e$ a further neighbour
$e',e''$ of degree at least $2$ and that $e$ has degree $2$ and
$e' \neq e''$. This gives the branching factor
$\tau(4-s,6-2s,6+3s)$. 
Sixth, the same case as fifth, but with $e'=e''$. In that case that branching
factor is $\tau(4-s,6-2s,6+2s)$. 
All the subcases together have
$\max\{\tau(4-s,4+3s,4+3s),\tau(4-s,5,7-s),\tau(4-s,6-2s,6+2s)\}$
as worst case.

In (e) the degrees of $d',d'',e',e''$ are at least $2$ and $e'\neq e''$
as otherwise case (d) would apply. Again one makes a case distinction:
The first case is that $d,d'$ have both degree $2$; here one assumes that
$e',e''$ have degree $3$ to minimise the weight-loss when removing $d,d'$.
Now the branching factor is $\tau(4-s,4+2s,6)$ or better, as when removing
$d',d''$, the neighbours $e',e''$ lose weight $1-s$ each.
The second case is as before, but $e'$ has degree $2$; then when removing
$a$ one can apply simplification rule {\bf S3} and remove $d',e'$ and a
further node as well. The branching factor is $\tau(4-s,4+3s,5+2s)$ or
better and this is better than $\tau(4-s,4+2s,6)$ from above.
The third case is that $d'$ has degree $3$.
Then the branching factor is at least
$\max\{\tau(4-s,5,7-s),\tau(4-s,6-2s,8-2s)\}$ where the two cases are
degree of $d''$ being $2$ versus $3$ and the degrees of $e',e''$
are assumed to be $3$ for worst-case weight loss.
In summary, the branching factor of this case is $\max\{\tau(4-s,4+2s,6),
\tau(4-s,5,7-s),\tau(4-s,6-2s,8-2s)\}$.

In summary the worst case of these branching rules is
$\max\{\tau(4-s,5,7-s)$, $\tau(4-s,4+3s,4+3s)$, $\tau(4-s,4+2s,6)$,
$\tau(4-s,6-2s,6+s),\tau(4-s,6-2s,8-2s)\}$.

\paraword{Four-Way Branching rule B4.1.}
Four-Way branching occurs only in the following situation:
\begin{verbatim}
B4.1.  d-|-d'-|-d''
\end{verbatim}
\ifx\longy\noy
The computation of an upper bound of
$\tau(6+2s,6+2s,6+2s,6+2s)$ for this rule has been moved to the appendix.
\fi
\def\movetotheappendixfourway{If $d'$
has a neighbour of degree $1$ in addition to $d',d''$,
a two-way branching would apply, thus one can assume that the
neighbour, if it exists, has at least degree $2$.
As the rule {\bf S4} does not apply, the nodes $d,d''$ have
also at least degree $2$. Furthermore, the node $d'$ has either
degree $2$ or degree $3$. Whenever $d'$ is removed from $V$
then two edges in the bisection cut are removed. Thus one has at least
one of the following branching factors (depending on whether $d'$
has degree $2$ or $3$):
$\tau(6+s,6+s,6+s),\tau(6+2s,6+2s,6+2s,6+2s)$ where the first
branching factor is better than the second due to three-way branching
instead of four-way branching. Note that for the three-way branching,
on the side of $d'$ of the bisection cut, only a node of value $s$ is
branched together with a gain of $6$ in each branch for the accounting
of two edges removed from the bisection cut. For the four way branching,
there are four choices of what happens with $d'$: Only removing $d'$
without adding an edge to $S$ versus adding exactly one of the edges
bording $d'$ into $S$.}
\ifx\longy\yesy
\movetotheappendixfourway
\fi

\begin{thm} \label{th:main}
The case distinction in the algorithm is complete.
\ifx\longy\noy
$($Please see the appendix for a proof.$)$
\fi
The runtime $O(1.2630^n)$ where $s$ is chosen
as $0.636$ giving the optimal entry in the following table.
\begin{center}
\begin{tabular}{|l|r|r|r|r|}
\hline
Rule & Formula & \ $s=0.6$ & 0.636 & 0.7 \\
\hline
{\bf B2.1, B2.2} & $\tau(3+s,4)$ & 1.2004 & \ 1.1993 & \ 1.1974 \\
     & $\tau(4-s,5-s)$ & 1.1958 & 1.1978 & 1.2015 \\
\hline
{\bf B3.1}, {\bf B3.2} $b$-side
  & $\tau(3+4s,3+4s,3+4s)$ & 1.2257 & 1.2192 & 1.2086 \\
  & $\tau(4,4+s,5+s)$ & 1.2644 & 1.2630 & 1.2606 \\
\hline
{\bf B3.2} $a$-side, {\bf B3.3}
  & $\tau(4-s,4+2s,6)$ & 1.2618 & 1.2615 & 1.2610 \\
  & $\tau(4-s,4+3s,4+3s)$ & 1.2544 & 1.2520 & 1.2478 \\
  & $\tau(4-s,5,7-s)$ & 1.2596 & 1.2612 & 1.2641 \\
  & $\tau(4-s,6-2s,6+s)$ & 1.2609 & 1.2630 & 1.669 \\
  & $\tau(4-s,6-2s,8-2s)$ & 1.2582 & 1.2617 & 1.2683 \\
\hline
{\bf B4.1} & $\tau(6\!+\!2s,6\!+\!2s,6\!+\!2s,6\!+\!2s)$ 
  & 1.2124 & 1.2030 & 1.2061 \\
\hline
Overall & Max of Above & 1.2644 & 1.2630 & 1.2683 \\
\hline
\end{tabular}
\end{center}
All branching factors in the table are strictly uprounded and
therefore the algorithm is in $O(1.2630^n)$ without additional
subexponential factors.
\end{thm}

\def\movetotheappendixtext{
\paravord{Proof.} The runtime follows from the fact that the branching
factors determine the size of the search tree and that every individual
instance of a rule can be carried out in time polynomially in the
number of involved nodes (without the recursive subcall).
The best-possible branching factor is then 1.2630 and obtained
by chosing $s$ optimally in the table given in the statement of
the theorem. So the main part is to verify that always one of the
rules applies, that is, that the preconditions where a rule can be applied
do not leave any case undone and form a complete case-distinction of
all possible cases. When several rules apply, the one applying first
is done.

\paraword{Completeness of Case-Distinctions.}
Now it is shown that as long as there are nodes in $V$,
one of the four cases in the algorithm applies.
If there are no edges in $B$, that is, if the previous bisection
cut has been completely branched or there has not yet any been made,
one checks first whether $(V,E)$ is connected. If not, then one
splits the graph into the independent components and solves each
of them independently. If $(V,E)$ is connected, one checks whether
there are more than $\kappa$ branching nodes. If not, simplification
rules {\bf S1} and {\bf S2} will solve the graph. If yes, then
one computes according to Corollary~\ref{co:monienpreis} the
corresponding bisection cut.

If there are edges in the bisection cut then one has to show that
one of the following rules applies:
Simplification rules {\bf S1}, {\bf S2}, {\bf S3}, {\bf S4} where the
first three also work when $B$ is empty and the fourth reduces
$B$ by one edge in the case that on one side of the bisection cut
is a node of degree $1$.

If none of these apply, there are the following
situations: {\bf B2.1} and {\bf B2.2} handle the situation that
the bisection cut has on one side a node $a$ of degree $3$ which
is connected with a node of degree $1$;
{\bf B3.1} handles the case
that both sides of the bisection cut are nodes of degree $2$,
{\bf B3.2} handles the case that on one side is a node of degree $2$
and on the other side is a node $a$ of degree $3$ which has all neighbours
having at least degree $2$; {\bf B3.3} handles the case that on both
sides of the bisection cut are directly nodes of degree $3$ which satisfy
that all neighbours have degree at least $2$.
{\bf B4.1} handles the case that a node has two neighbours in the bisection
cut which, by accounting, gives a better branching factors as the others,
even if one needs to do four-way branching.~\qed }

\ifx\longy\yesy
\movetotheappendixtext
\fi

\section{Conclusion}

\noindent
For various papers, the subcase of cubic graphs when solving the
maximum induced matching problem was only treated by invoking the
line graph argument of Cameron \cite{C89} which gives $O(1.3139^n)$.
The present polynomial space algorithm gives an overall runtime of
$O(1.2630^n)$; better algorithms use exponential space \cite{FH06,KK20}.
One of the ideas is not to split the graph using the Monien-Preis method
such that the number of nodes is balanced, but such that the number of
degree $3$ nodes is balanced; for evaluating in this situation the
algorithm, one looks at the branching numbers from both sides of the
bisection cuts and budgets for each edge branched of the bisection cut
an amortised gain of $3$ degree $3$ nodes which go to the other side;
all branching rules considered remove at least one edge of the bisection
cut.

\paraword{Acknowledgments.} The authors would like to thank Henning Fernau
and others for pointing out the work of Fomin and H\o ie \cite{FH06} and
the importance of pathwidth for exponential space algorithms to them.

\ifx\longy\yesy
\end{document}